\documentclass[aps, twocolumn, superscriptaddress]{revtex4} 

\usepackage[switch]{lineno}
\usepackage{amsmath}
\usepackage{amssymb}
\usepackage{empheq}
\usepackage[parfill]{parskip}
\usepackage[active]{srcltx}
\usepackage{color}
\usepackage{array}
\usepackage{booktabs}
\usepackage{amsfonts}
\usepackage{dsfont}
\usepackage{graphicx}

\begin{document}

\setlength{\parindent}{0.5cm}

\title{Pinning in a system of swarmalators}

\author{Gourab Kumar Sar}
\email{mr.gksar@gmail.com}
\affiliation{Physics and Applied Mathematics Unit, Indian Statistical Institute, 203 B. T. Road, Kolkata 700108, India}

\author{Dibakar Ghosh}
\email{diba.ghosh@gmail.com}
\affiliation{Physics and Applied Mathematics Unit, Indian Statistical Institute, 203 B. T. Road, Kolkata 700108, India}

\author{Kevin O'Keeffe}
%\email{kevin.p.okeeffe@gmail.com}
\affiliation{Senseable City Lab, Massachusetts Institute of Technology, Cambridge, MA 02139}

\begin{abstract}
\hspace{1 cm}  (Received XX MONTH XXX; accepted XX MONTH XXXX; published XX MONTH XX)\\ 

We study a population of swarmalators (swarming/mobile oscillators) which run on a ring and are subject to random pinning. The pinning represents the tendency of particles to stick to defects in the underlying medium which competes with the tendency to sync / swarm. The result is rich collective behavior. A highlight is low dimensional chaos which in systems of ordinary, Kuramoto-type oscillators is uncommon. Some of the states (the phase wave and split phase wave) resemble those seen in systems of Janus matchsticks or Japanese tree frogs. The others (such as the sync and unsteady states) may be observable in systems of vinegar eels, electrorotated Quincke rollers, or other swarmalators moving in disordered environments. \\
\noindent \\
DOI: XXXXXXX
\end{abstract}

\maketitle
%%%%%%%%%%%%%%%%%%%%%%%%%%%%%%%%%%%%%
\section{Introduction}
Synchronization (self-organization in time) and swarming (self-organization in space) are universal phenomena that co-occur in driven colloids \cite{yan2012linking,hwang2020cooperative,zhang2020reconfigurable,bricard2015emergent,zhang2021persistence,manna2021chemical,li2018spatiotemporal,chaudhary2014reconfigurable}, embryonic cells \cite{tan2021development,petrungaro2019information}, biological microswimmers  \cite{yang2008cooperation,riedel2005self,quillen2021metachronal,quillen2022fluid,peshkov2022synchronized,tamm1975role,verberck2022wavy,belovs2017synchronized},  and robotic swarms \cite{barcis2019robots,barcis2020sandsbots,schilcher2021swarmalators,monaco2020cognitive,hadzic2022bayesian,rinner2021multidrone,gardi2022microrobot,origane2022wave}. Yet a theoretical understanding of the interplay between sync and swarming is lacking. Much is known about sync \cite{pikovsky2001universal,winfree1980geometry,kuramoto2003chemical} and swarming \cite{toner1995long,vicsek1995novel} acting independently of each other, yet little is known about their \textit{interaction} -- an effect which some have dubbed `swarmalation' \cite{verberck2022wavy,riedl2022synchronization}, since it combines \underline{swarm}ing in space with oscil\underline{lation} in time. %\footnote{Actually, the entities `swarmalators', short for swarming oscillators, was introduced first, and then the abstract noun `swarmalation` followed by completing the parallel: If oscillators cause oscillation, then swarmalators cause swarmalation}.

Swarmalation, as theoretical field, began about 15 years ago when Tanaka et al. introduced a model of chemotactic oscillators \cite{tanaka2007general,iwasa2010hierarchical,iwasa2012various,iwasa2017mechanism}. Later O'Keeffe et al. studied a generalized Kuramoto model of `swarmalators' \cite{o2017oscillators} which triggered a new wave of work \cite{hong2018active,lee2021collective,o2018ring,o2019review,o2022collective,o2022swarmalators,jimenez2020oscillatory,schilcher2021swarmalators,japon2022intercellular,sar2022swarmalators,yoon2022sync,ha2021mean,ha2019emergent,degond2022topological,sar2022dynamics,mclennan2020emergent,blum2022swarmalators,monaco2019cognitive,monaco2020cognitive,schultz2021analyzing,hadzic2022bayesian,ceron2022diverse}. Even so, the sub-field is still very much in its infancy, with little known in the way of analysis \cite{yoon2022sync,o2022swarmalators}, and even less known in the way of swarmalator phenomenology.

This paper sets out to explore this uncharted terrain with a case-study of swarmalators subject to random pinning. Random pinning is a well studied topic in nonlinear dynamics and statistical physics. It refers to the tendency of a substance to stick to the impurities of the underlying medium requiring forcing to produce flow. A classic example is charge density waves \cite{gor2012charge,gruner1985charge} which have been analyzed using phases oscillators \cite{strogatz1989collective,strogatz1988simple}. Here, the phase $\theta_i$ corresponds to the phase of the density wave  at a (fixed) position $x_i$ in the underlying lattice $\rho(x,t) = \rho_0 \sin( kx - \theta(x,t)$). At a critical forcing, the phases depin from their preferred values $\theta_i^*$ producing interesting collective effects. 

Charge density waves partly inspired our study of pinned swarmalators. We wondered: what might happen if the lattice sites $x_i$ themselves begin to vibrate and interact with the density waves' phases $\theta_i$?

%A thought experiment about charge densities waves in fact inspired our study of pinned swarmalators. We ask: what might happen if the lattice sites $x_i$ themselves begin to vibrate and interact with the density waves' phases $\theta_i$?

%Charge density waves provide a motivation for our study of pinned swarmalators. We ask: what might happen if the lattice sites $x_i$ themselves begin to vibrate and interact with the density waves' phases $\theta_i$?

A clue comes from a recent work on a pair magnetic domains walls \cite{hrabec2018velocity,haltz2021domain}, entities with bright futures as memory devices in next generation  spintronics \cite{Foerster2016,wolf2001spintronics,ohno2010window}. When sufficiently forced, the center of mass $x_i$ and the magnetic dipole vector $\theta_i$ of the walls unlock and begin to interact producing rich spatiotemporal dynamics (see Fig 2 and the lissajous curves in Supp. Figs 6,7 in \cite{hrabec2018velocity}). Just $N = 2$ swarmalators produce these dynamics. We expect $N \gg 1$ swarmalators -- the regime we are here interested in -- to produce even richer dynamics. 

We search for such dynamics using a simple model of swarmalators which move on a 1D ring. The hope is that a bare-bones, simple model might capture phenomena with some universality. We find diverse collective behavior: a family of phase waves, ``flying saucers'' where swarmalators form traveling loops in $(x,\theta)$ space, quasi-periodicity, and chaos. The unsteady behaviors do not occur in systems of pinned oscillators; they are exclusive to systems of pinned swarmalators. 

We hope our work inspires more studies of swarmalators with random pinning. As we show, the model is tractable and could thus be used as a testbed for general studies of swarmalation in disordered media.

%%%%%%%%%%%%%%%%%%%%%%%%%%%%%%%%%%%%%%%%%%%%%%%%%%%%%%%%%%%%%%%%%%%%%%%%%%%%%%%%%%%
\begin{figure}[htpb!]
\centering
\includegraphics[width = \columnwidth]{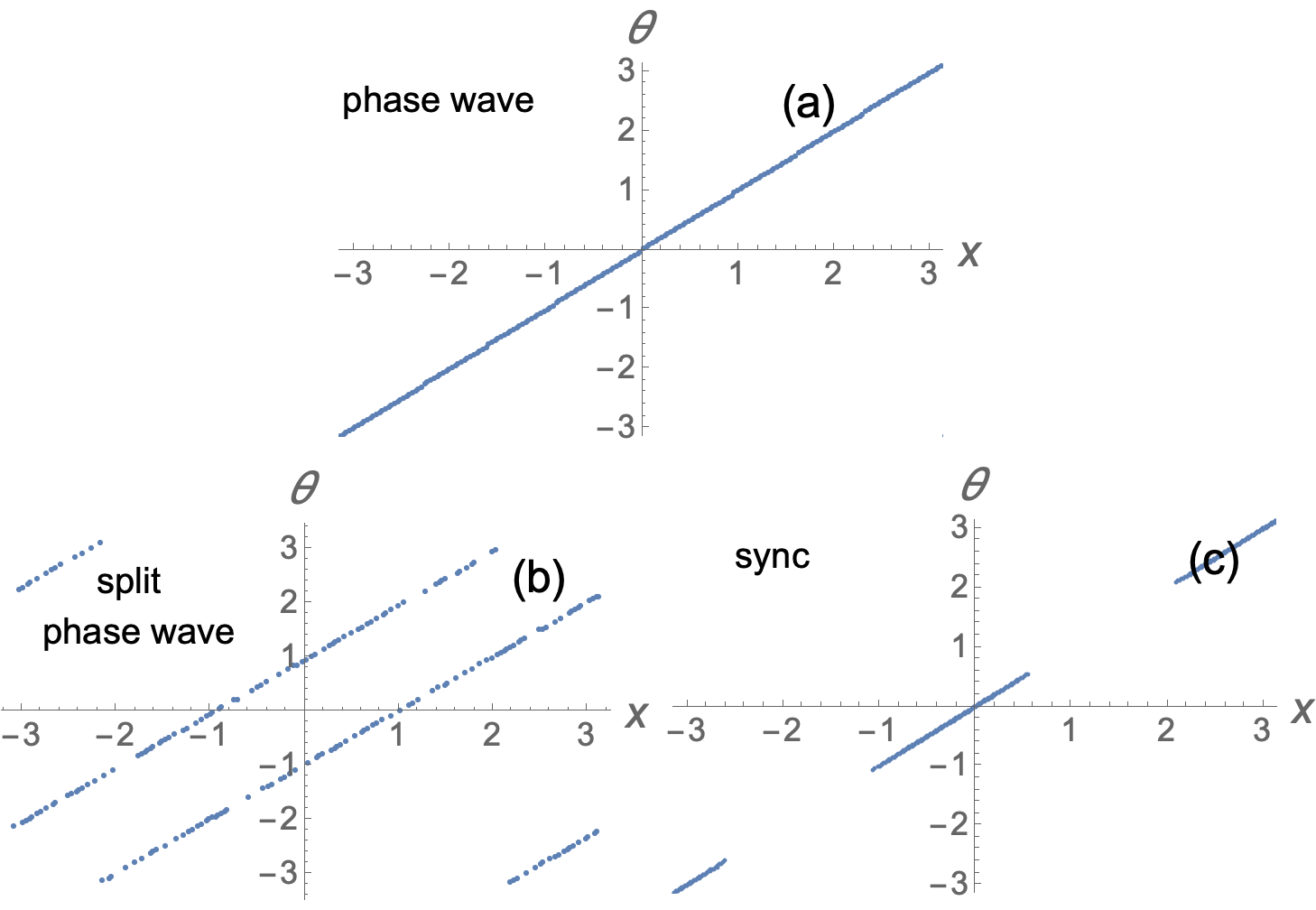}
\caption{\textbf{Collective states in undriven limit $E = 0$.} (a) Pinned state $K = 1$, (b) split phase wave $K = -2$, (c) sync state $K = 3$. In each panel, $(dt,T,N) = (0.1, 100, 200)$ and initial $x_i, \theta_i$ were drawn randomly from $[-\pi,\pi]$.}
\label{states-undriven}
\end{figure}
\section{Model}
We study a pair of Kuramoto-like models
\begin{align}
    \dot{x_i} &= E -b \sin(x_i - \alpha_i) + \frac{K}{N} \sum_j^N \sin(x_j - x_i) \cos(\theta_j - \theta_i), \label{eom-x} \\
    \dot{\theta_i} &= E -  b \sin(\theta_i - \beta_i) + \frac{K}{N} \sum_j^N \sin(\theta_j - \theta_i ) \cos(x_j - x_i ), \label{eom-theta}
\end{align}
\noindent
where $(x_i, \theta_i) \in (\mathbb{S}^1, \mathbb{S}^1)$ are the position and phase of the $i$-th swarmalator. $E$ represents external forcing, $K$ the inter-element couplings, $b$ the strength of the pinning, and $\alpha_i, \beta_i$ the pinning locations. We confined the motion to a 1D ring mainly for simplicity. That said, there are real-world swarmalators which move around 1D rings; vinegar eels move along the ring-like edges of 2D disks, Japanese tree frogs on the edges of paddy fields \cite{aihara2014spatio}, and coupled magnetic bilayers, above the walker regime, oscillate along on a 1D axis. We pin the swarmalators' positions uniformly around the circle  $\alpha_i =  2\pi i/N$ and also set the phase pinning equal to the space pinning $\beta_i = \alpha_i$. These choices simplify the analysis, and also enforce that in the the pinned state ($K = 0$) the swarmalators form a phase wave around the circle $x_i = \theta_i$ -- a natural base state which occurs commonly in Nature (see Discussion).  Finally, we set $b = 1$ without loss of generality, which leaves a two-parameter model $(K,E)$.

We simplify our model by converting the trigonometric functions to complex exponentials,
\begin{align}
    \dot{x_i} &= E - \sin(x_i - \alpha_i) + \frac{K}{2}  S_+ \sin( \Phi_+ - (x_i + \theta_i) ) \nonumber \\ 
    \ &+ \frac{K}{2}  S_- \sin( \Phi_- - (x_i - \theta_i) ), \label{eq3}\\
    \dot{\theta_i} &= E - \sin(\theta_i - \beta_i) + \frac{K}{2} S_+ \sin( \Phi_+ - (x_i + \theta_i) ) \nonumber \\
    & - \frac{K}{2}  S_- \sin( \Phi_- - (x_i - \theta_i) ), \label{eq4}
\end{align}
\noindent
where 
\begin{equation}
    W_{\pm} = S_{\pm} e^{\Tilde{i} \Phi_{\pm}} =  \frac{1}{N} \sum_j e^{\Tilde{i}(x_j \pm \theta_j)} \hspace{10pt} (\Tilde{i} = \sqrt{-1})\label{order-par}
\end{equation}
\noindent
are order parameters which measure the system's space-phase order. When positions and phases are perfectly correlated $x_i = \pm \theta_i + C$, they are maximal $S_{\mp} = 1$. When $x_i, \theta_i$ are uncorrelated, they are minimal $S_{\pm} = 0$. 

Now we begin our analysis.
%%%%%%%%%%%%%%%%%%%%%%%%%%%%%%%%%%%%%%%%%%%%%%%%%%%%%%%%%%%%%%%%%%%%%%%%%%%%%%%%%%%%%%%%%%%%%%
\section{No driving $E = 0$}
To warm up, we study the undriven system $E = 0$ (which leaves just one active parameter $K$). Numerics reveals three collective steady states as we vary $K$: the \textit{phase wave}, \textit{split phase wave}, and \textit{sync state}. The first three panels of Supplementary Movies 1\&2 show the evolution to these states.

\begin{figure}[t]
\centering
\includegraphics[width = \columnwidth]{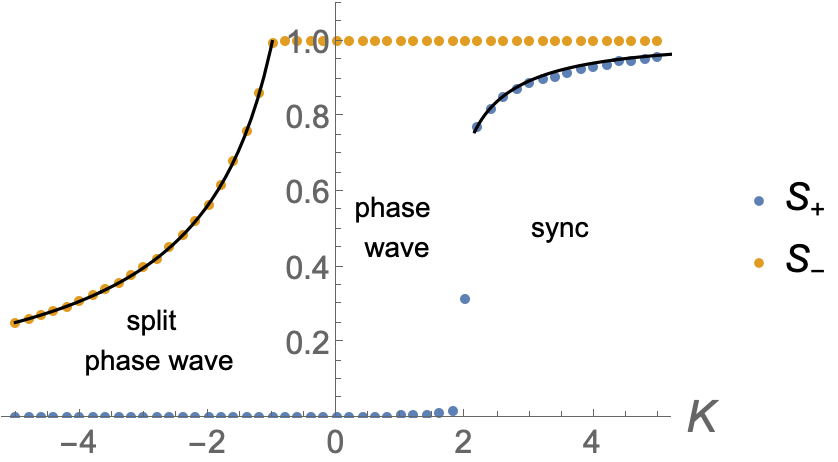}
\caption{\textbf{Order parameters in undriven limit $E = 0$.}  Colored dots show simulation results. The thick black curve on the LHS shows theoretical prediction Eq.~\eqref{sm}, the one on the RHS the numerical solution of Eqs.~\eqref{s1} and \eqref{s2}. Hysteresis exists in the small region $K \in [0.1926, 2]$, which is plotted more clearly in Fig.~\ref{sync-k-sp}. Simulation parameters: $(dt, T, N) = (0.25, 200, 400)$.}
\label{order-par-undriven}
\end{figure}

\textbf{Phase wave}. When the pinning dominates the synchronizing $ |K| \lessapprox 1$ (or in dimensionful units $|K| \lessapprox b$), the swarmalators' positions and phases stay pinned to their preferred values $ x_i = \theta_i = \alpha_i = 2\pi i / N$. As mentioned, in the $(x,\theta)$ plane, this constitutes a phase wave $x_i = \theta_i$ (Fig.~\ref{states-undriven}(a)) which implies $(S_+, S_-) = (0,1)$ (Fig.~\ref{order-par-undriven}). The stability of this state in the no pinning limit $b = 0$ was calculated previously by linearizing the $(\dot{x}_i, \dot{\theta}_i)$ equations and exploiting the block structure of the Jacobian $\mathcal{J}$ \cite{o2022collective}. Adapting this calculation when pinning is turned on $b = 1$ yields three distinct eigenvalues:
%\begin{align}
%    \lambda_0 &= -b \label{4}\\
%    \lambda_1 &= -\frac{1}{2}(2b + J + K) \label{5}\\
%    \lambda_2 &=  -\frac{1}{8} \left(8 b + K + J \pm \sqrt{K^2+34K+J^2}\right) \label{6}
%\end{align}
\begin{align}
    \lambda_0 &= -1, \\
    \lambda_1 &= -1 - K, \label{2} \\
    \lambda_2 &= \frac{1}{2} (-2 + K), \label{3}
\end{align}
with multiplicities $N-1$, $N-1$, and $2$, respectively. These imply the pinned state is stable for
\begin{equation}
-1 \leq K \leq 2 \label{s5}.
\end{equation}
This holds true for all finite $N$. \\

%In the phase wave, neighboring swarmalators positions and phases are differ by $2 \pi / N$. When the coupling becomes sufficiently negative $K < -1$, however, swarmalators want to desynchronize, or put another way, want to maximize their phase differences. The result is that

\textbf{Split phase wave}. When $K < -1$, the phase wave splits into two phase waves (Fig.~\ref{states-undriven}(b)) separated by a distance $\Delta(K)$ from the pinned state:
\begin{align}
    x_i &= \alpha_i + (-1)^i \Delta, \label{t1} \\
    \theta_i &= \alpha_i + (-1)^{i-1} \Delta. \label{t2} 
\end{align}
The order parameter $S_+$ remains zero, but now $S_-$ obeys $0 \leq S_-(K) < 1$ (Fig.~\ref{order-par-undriven}) and must be solved for self consistently. We insert Eqs.~\eqref{t1} and \eqref{t2} into the $\dot{x}_i$ equation and apply the fixed point condition $\dot{x}_i =0$,
\begin{align}
    \dot{x}_i &= \sin \Big( (-1)^i \Delta \Big) +  \frac{K}{2} S_- \sin(2 \Delta (-1)^{i-1}) = 0. \label{t3}%\nonumber \\\
%    0 &= \sin \Big( (-1)^i \Delta  \Big) + \frac{K}{2} S_- \sin(2 \Delta (-1)^{i-1}) 
\end{align}
(The $\dot{\theta}_i$ equation is a duplicate so we omit it). Equation~\eqref{t3} is in fact independent of $i$. When $i$ is even, the first term is positive, the second negative. When $i$ is odd the reverse happens. Since the RHS is zero, the sign flipping doesn't matter so we set $i=0$ without loss of generality and rearrange terms to obtain
\begin{align}
   S_- = - \frac{2}{K} \frac{\sin \Delta}{ \sin 2 \Delta}. \label{t4}
\end{align}
Next we plug Eqs.~\eqref{t1} and \eqref{t2} into the definition for $S_-$:
\begin{align}
   S_- &= \frac{1}{N} \sum_j e^{\Tilde{i}(x_j - \theta_j)}, 
\end{align}
which gives us,
\begin{align}
   S_- &= \cos 2\Delta \label{t5}.
\end{align}
Equations~\eqref{t4} and~\eqref{t5} comprise a pair of simultaneous equations for $(\Delta,S_-)$. To solve them we eliminate $\Delta$ by writing $\sin \Delta, \sin 2\Delta$ in terms of $\cos 2 \Delta = S_-$. Using the identity $\sin^2 2 \Delta + \cos^2 2\Delta = 1$, we find $\sin 2\Delta = \pm \sqrt{1 - \cos^2 2\Delta} = \pm \sqrt{1 - S_-^2}$. Using the identity, $\cos 2\Delta = 1 - 2\sin^2 \Delta$, we find $\sin \Delta = \pm \sqrt{(1 - \cos 2\Delta)/2} = \pm \sqrt{(1 - S_-)/2}$. Plugging these into Eq.~\eqref{t4} gives 
\begin{equation}
    S_- = \pm \frac{2}{K} \sqrt{\frac{1-S_-}{2(1-S_-^2)}}.
\end{equation}
Cleaning this up yields a cubic in $S_-$
\begin{equation}
   K^2 S_-^3 + K^2 S_-^2 - 2 = 0.
\end{equation}
The unique real root is
\begin{align}
S_- &= -1 + \frac{K^2}{\sqrt[3]{-K^6+27 K^4+3 \sqrt{81 K^8-6 K^{10}}}} \nonumber \\
    & \; +\frac{\sqrt[3]{-K^6+27 K^4+3 \sqrt{81 K^8-6 K^{10}}}}{K^2}. \label{sm}
\end{align}
This \textit{exists} for all $K$, but becomes \textit{stable} at $K = -1$ via a transcritical bifurcation when it intersects with the $S_- = 1$ branch (Fig.~\ref{order-par-undriven}). As $K \rightarrow -\infty$, $S_{-}$ declines from $1$ to $0$, and the `split gap' $\Delta$ increases from $0$ to $\pi/4$.
\begin{figure}[t!]
\centering
\includegraphics[width = 0.85 \columnwidth]{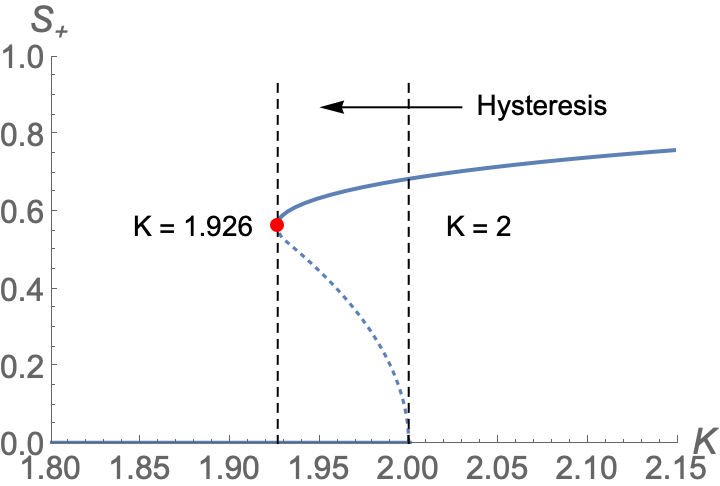}
\caption{ \textbf{Hysteresis} in $S_+(K)$ for $E = 0$. Curve is drawn by solving Eqs.~\eqref{s1} and~\eqref{s2} numerically. It shows hysteretic transition from the phase wave to the sync state.}
\label{sync-k-sp}
\end{figure}

\textbf{Sync state}. We have sussed out the split phase wave on $K < -1$ and the phase wave on $-1 \leq K \leq 2$. Now we jump to remaining region $K > 2$, where the sync force dominates the pinning force, so to speak. Here swarmalators form synchronous clusters (Fig.~\ref{states-undriven}(c)). Two clusters form when the initial conditions $x_i, \theta_i$ are drawn uniformly at random from $[-\pi, \pi]$, while one cluster forms when $x_i, \theta_i$ are drawn from $ [0, \pi]$ (the single cluster state is not plotted). Figure~\ref{order-par-undriven} shows $S_-$ stays at unity while $S_+$ bifurcates from $0$ at $K_c = 2$.

We again solve for $S_+(K)$ self consistently. This time we move to coordinates
\begin{align}
    \xi & := x + \theta, \\
    \eta & := x - \theta.
\end{align}
The governing ODEs in this frame are
\begin{align}
    \dot{\xi}_i = -2 \cos \frac{\eta_i}{2} \sin \Big( \frac{\xi_i}{2} - \alpha_i \Big) + K S_+ \sin (\Phi_+ - \xi_i), \\
    \dot{\eta}_i = -2 \sin \frac{\eta_i}{2} \cos \Big( \frac{\xi_i}{2} - \alpha_i \Big) + K S_- \sin (\Phi_- - \eta_i).
\end{align}
Next we search for fixed points. First we simplify by setting $S_- = 1$ and $\eta_i = \Phi_- = 0$ (these ansatzes were suggested by numerics). The $\dot{\eta}_i$ equation reduces to $0 = 0$. The $\dot{\xi}_i$ equation reads
\begin{align}
    -& 2 \sin \Big( \frac{\xi}{2} - \alpha \Big) + K S_+ \sin (\Phi_+ - \xi)  = 0, \label{s1} \\
    & S_+ e^{\Tilde{i} \Phi_+} = \frac{1}{2\pi}\int_{0}^{2\pi} e^{\Tilde{i} \xi(\alpha)} d \alpha , \label{s2} 
\end{align}
where Eq.~\eqref{s2} requotes the definition of $S_+$ in the $N \rightarrow \infty$ limit. We must solve Eq.~\eqref{s1} for the fixed points $\xi^*(\alpha)$ then plug them into Eq.~\eqref{s2} to find $S_+$. First we set $\Phi_+ = 0$ without loss of generality. Then by applying various trig identities to Eq.~\eqref{s1} we arrive 4-th order polynomial in $\cos \xi^*$ and plug the roots into Eq.~\eqref{s2}, which when $\Phi_+ = 0$ reads $S_+ = \int \cos(\xi^*(\alpha)) d \alpha$. Unfortunately, we were unable to do the integral over $\alpha$ analytically. So instead we approximated it numerically. Figure~\ref{sync-k-sp} shows it matches simulation. Notice there is a bistable region $K \in [1.926, 2]$. 
\begin{figure}[t!]
\centering
\includegraphics[width = \columnwidth]{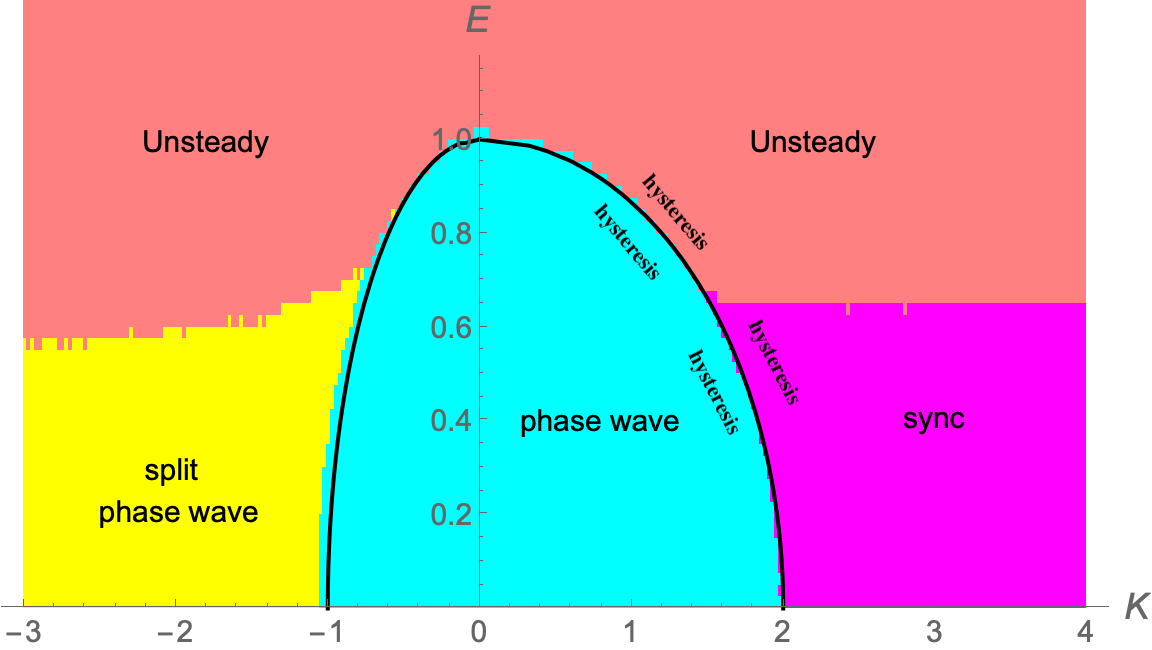}
\caption{\textbf{Bifurcation diagram in $(K,E)$ plane}. Black curves denote theoretical predictions. Colored region were found by computing $(S_+, S_-, \langle V \rangle)$ over the population over a $(K,E)$ mesh and applying various conditions (see main text). Hysteresis boundaries were not computed; the text just indicates their approximate locations. Here $(dt, N, T) = (0.1, 200, 200)$. The first $90\%$ of simulation data were discarded as transients. The jagged boundaries are due to finite effects in the $(K,E)$ mesh.}
\label{bif-diagram}
\end{figure}

This completes our study of the undriven system $E = 0$. To recap, we solved for the order parameters in each collective states: $S_-(K)$ in the split phase wave, $S_+(K)$ in the sync state, and the trivial result $(S_+, S_-) = (0,1)$ in the phase wave. We also determined the stability of the phase wave for all finite $N$. From a dynamic point of view, the discovered states were tame; in each, swarmalators were ultimately stationary in space and phase. When external driving is turned on $E >0$, however, the dynamics become richer. 

%%%%%%%%%%%%%%%%%%%%%%%%%%%%%%%%%%%%%%%%%%
\section{Driving $E > 0 $}
The phase wave, split phase wave, and sync state persist for small amount of driving $E > 0$. This is a good news; it means we can use our earlier $E = 0$ analyses of these states as springboards to the $E > 0$ regime. 

To help guide our analysis, we first numerically constructed the bifurcation diagram in the $(K,E)$ plane. We computed $(S_+, S_-, \langle V \rangle)$ at each point in a uniform $(E_i, K_i)$ mesh, where $\langle V \rangle := N^{-1} \sum_i \sqrt{ v_{x,i}^2 + v_{\theta,i}^2 } $ is the mean velocity. Then we divided the plane into four regions according to these conditions:
\begin{align}
    & \text{Unsteady}: \langle V \rangle  > 0. \nonumber \\
    & \text{Phase wave}: \{\langle V \rangle  = 0\} \cap \{S_{+} < 0.1 \} \cap \{S_{-} > 0.99\}. \nonumber\\
    & \text{Split phase wave}: \{\langle V \rangle  = 0\} \cap \{S_{+} < 0.1\} \cap \{S_{-} < 0.99\}. \nonumber \\
    & \text{Sync}: \{\langle V \rangle  = 0\} \cap \{S_{+} > 0.1\} . \nonumber
\end{align}
Figure~\ref{bif-diagram} shows the result. Our analysis will be a walk through each of the colored regions / states, so keep this map in mind as we go.

\textbf{Split phase wave}. Here the analysis is essentially a repeat of the $E = 0$ case. The fixed points become
\begin{align}
    x_i &= \alpha_i + \sin^{-1}(E) + (-1)^i \Delta ,\\
    \theta_i &= \beta_i + \sin^{-1}(E) + (-1)^{i-1} \Delta .
\end{align}
the resultant cubic is
\begin{equation}
   K^2 S_-^3 + K^2 S_-^2 - 2 + 2E^2 = 0 ,
   \label{eq28}
\end{equation}
whose unique real solution is
\begin{align}
    S_- &= -\frac{1}{3} + \frac{K^4 + (\Gamma_1 + 3 \sqrt{3} \sqrt{ \Gamma_2} )^{2/3}}{3 K^2 (\Gamma_1 + 3 \sqrt{3} \sqrt{ \Gamma_2})^{1/3} } , \label{smE} 
\end{align}
where
\begin{align}
    \Gamma_1 & := -27 \left(E^2-1\right) K^4-K^6 , \\
    \Gamma_2 & := \left(E^2-1\right) K^8 \left(27 E^2+2 K^2-27\right) .
\end{align}
We can use Eq.~\eqref{smE} to study the bifurcations of the split phase wave. The trick is to peel off the critical $(E,K)$ boundaries by evaluating Eq.~\eqref{smE} at extremal values of $S_-$ (these  values are $0,1$ since $0 \leq S_{\pm} \leq 1$). First we set $S_- = 1$ (which corresponds to the bifurcation to the phase wave) and solve to find
\begin{align}
    E_c(K) = \sqrt{1 - K^2}. \label{E_c}
\end{align}
This is an existence condition; it doesn't say anything about stability. Simulations show however that it coincides with the stability boundary as shown in Fig.~\ref{bif-diagram} (the black curve between the yellow and cyan regions). 

\begin{figure}[t!]
\centering
\includegraphics[width = \columnwidth]{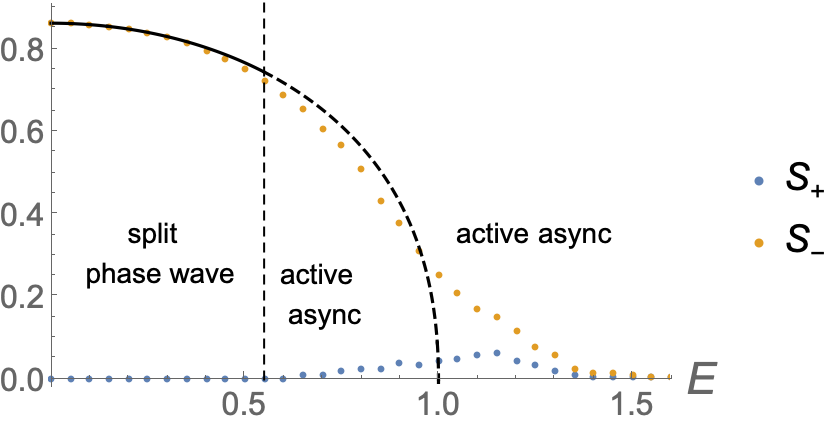}
\caption{ \textbf{Order parameter with driving}. $S_{-}$ versus $E$ for $K = -1.2$. Colored dots show simulation results. The black curve shows theoretical prediction Eq.~\eqref{smE}. Notice $S_- < 0.99$ exists for $E \in [0,1]$, but is stable for $ E < E^{**} \approx 0.55$. For $E > E^{**}$ the active async state is realized. Also notice $S_+$ becomes non-zero for $E > E^{**}$. Here $(dt,T,N) = (0.25, 200, 200)$ and data are averaged over last 10 \% of data collected.}
\label{Sm-E}
\end{figure}

Now we target the other extreme $S_- = 0$ which has existence condition
\begin{align}
    E^*(K) = 1 . \label{Estar}
\end{align}
This time existence and stability boundaries do \textit{not} coincide; the split phase wave loses stability at some $E^{**} < E^*$. One can see this in Fig.~\ref{bif-diagram}, where the boundary between the yellow and light red regions is clearly not $E^* = 1$. Figure~\ref{Sm-E} confirms this picture by plotting $S_-(E)$ for $K = -1.2$. For $ E < E^{**} \approx 0.55$, the prediction matches simulation. For $E > E^{**}$ however, theory and numerics start to diverge. Moreover, $S_+$ jumps from zero which certifies the transition to a new state  we call \textit{active async}.

\begin{figure}[htpb!]
\includegraphics[width = \columnwidth]{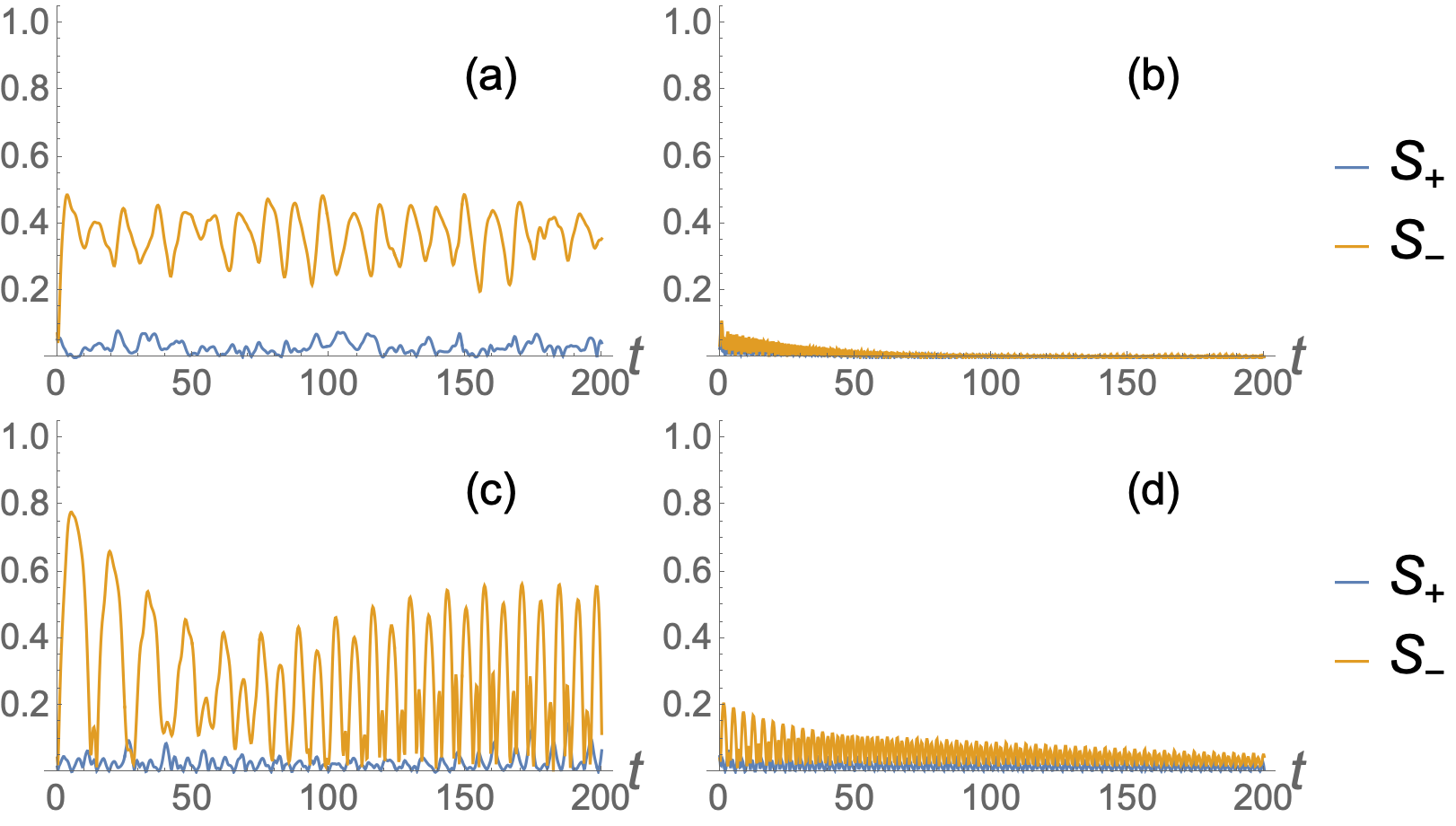}
\caption{\textbf{Active async}. Order parameters in active async state. (a) $(K,E) = (-1.5, 0.8)$, (b) $(K,E) = (-1.5, 3)$,  (c) $(K,E) = (-0.1, 1.1)$, (d) $(K,E) = (-0.1, 2)$. Simulation parameters: $(dt,T,N) = (0.25, 200, 300)$. }
\label{bursty-async}
\end{figure}

\textbf{Active async}. This state is best understood by watching Supplementary Movies 1\&2. Starting in the split phase wave at $K = -1.5$ (the yellow region of Fig.~\ref{bif-diagram}), we move $E$ just past $E^{**}$, and find swarmalators exhibit fast/slow dynamics. They move slowly about the `ghost' of the split phase wave, make sudden cycles in position $x$ and phase $\theta$, then return to the ghost. This burstiness manifests in $S_-$ as irregular oscillations (Fig.~\ref{bursty-async}(a)). As the driving becomes more extreme $E > E^{**}$, the slow-fast mechanism dies out and all swarmalator are pushed around $x,\theta$ at about the same speed resulting in $S_{\pm} \approx 0$ (Fig.~\ref{bursty-async}(b)). The bottom row of Fig.~\ref{bursty-async} shows the transition to async from the phase wave state (i.e. starting at $K = -0.1$; the cyan region in Fig.~\ref{bif-diagram}). The picture is qualitatively the same: large amplitude oscillations which shrink as the driving grows.

Unfortunately, we were unable to calculate the stability of this unsteady state. Numerics however show state dies for $K > 0$, the regime we now explore.

\textbf{Phase wave}. We begin the $K > 0$ exploration with the phase wave. This time however, we analyse in the $N \rightarrow \infty$ limit. The governing equations become
\begin{align}
    \dot{x}_{\alpha} &= E - \sin(x_{\alpha} - \alpha) + \frac{K}{2}  S_+ \sin( \Phi_+ - (x_{\alpha} + \theta_{\alpha}) ) \nonumber \\ 
    \ &+ \frac{K}{2}  S_- \sin( \Phi_- - (x_{\alpha} - \theta_{\alpha}) ) , \\
    \dot{\theta}_{\alpha} &= E - \sin(\theta_{\alpha} - \alpha) + \frac{K}{2} S_+ \sin( \Phi_+ - (x_{\alpha} + \theta_{\alpha}) ) \nonumber \\
    & - \frac{K}{2}  S_- \sin( \Phi_- - (x_{\alpha} - \theta_{\alpha}) ) ,
\end{align}
where $S_{\pm}$ are
\begin{equation}
    S_{\pm} e^{\Tilde{i} \Phi_{\pm}} =  \frac{1}{2\pi} \int_{0}^{2\pi} e^{\Tilde{i}(x_{\alpha} \pm \theta_{\alpha})} \, d\alpha ,
    \label{order-par-inf}
\end{equation}
and $x: \alpha \mapsto x_{\alpha}$ and $\theta: \alpha \mapsto \theta_{\alpha}$ should be thought as self-maps of the unit circle. The fixed points take form
\begin{equation}
    x_{\alpha} = \theta_{\alpha} = \alpha + \sin^{-1}(E) .\\
\end{equation}
Now, our goal is to compute the local stability of the phase wave. To get there, we follow the strategy in \cite{strogatz1989collective} which is to diagonalize the second variation of the potential function
\begin{align}
    V(x_{\alpha},\theta_{\alpha}) = -E \int_0^{2\pi} x_{\alpha} \, d \alpha \; -E \int_0^{2\pi} \theta_{\alpha} \, d \alpha \nonumber \\ - \int_0^{2\pi} \cos(\alpha - x_{\alpha}) \, d \alpha - \int_0^{2\pi} \cos(\alpha - \theta_{\alpha}) \, d \alpha  \nonumber \\ - \frac{K}{2\pi} \int_0^{2\pi} \int_0^{2\pi} \cos(x_{\beta} - x_{\alpha}) \cos(\theta_{\beta}-\theta_{\alpha}) \, d\alpha d \beta.
    \label{pot}
\end{align}
Simplification of Eq.~\eqref{pot} using $x_{\alpha} = \theta_{\alpha}$ yields
\begin{align}
    V(x_{\alpha}) &= -2E \int_0^{2\pi} x_{\alpha} \, d \alpha \; - 2\int_0^{2\pi} \cos(\alpha - x_{\alpha}) \, d \alpha \nonumber \\ &- \frac{K}{4\pi} \int_0^{2\pi} \int_0^{2\pi} \cos^2(x_{\beta} - x_{\alpha}) \, d\alpha d \beta.
    \label{pot1}
\end{align}
Let $\eta:\alpha \mapsto \eta_{\alpha}$ denote a perturbation about the phase wave state so that
\begin{equation}
    x_{\alpha}(\epsilon) = \alpha + \sin^{-1}(E) + \epsilon \eta_{\alpha} ,
    \label{perturb}
\end{equation}
where $\epsilon$ is small. Plugging Eq.~\eqref{perturb} into Eq.~\eqref{pot1} yields
\begin{align}
    V(x_{\alpha}&(\epsilon)) = -2E \int_0^{2\pi} (\alpha + \sin^{-1}(E) + \epsilon \eta_{\alpha}) \, d \alpha \; \nonumber \\ &- 2\int_0^{2\pi} \cos(\sin^{-1}(E) + \epsilon \eta_{\alpha}) \, d \alpha \nonumber \\ &- \frac{K}{4\pi} \int_0^{2\pi} \int_0^{2\pi} \cos^2(\beta - \alpha +\epsilon \eta_{\beta} - \epsilon \eta_{\alpha}) \, d\alpha d \beta.
    \label{pot2}
\end{align}
The second variation of $V$ is a quadratic form
\begin{equation}
    \Gamma(\eta) = \frac{d^2}{d \epsilon^2} V(x_{\alpha}(\epsilon))\Big |_{\epsilon=0}.
\end{equation}
Differentiation yields
\begin{align}
    \Gamma(\eta) &= 2 \sqrt{1-E^2} \int_{0}^{2\pi} \eta^2_{\alpha} \, d\alpha \nonumber \\
    &+ \frac{K}{2\pi} \int_{0}^{2\pi} \int_{0}^{2\pi} (\eta_{\beta} - \eta_{\alpha})^2 \cos(2\beta - 2 \alpha) \, d \alpha d \beta .
    \label{sec-var}
\end{align}
We can simplify the second term on the RHS of Eq.~\eqref{sec-var} by expanding the term $(\eta_{\beta} - \eta_{\alpha})^2$. Which gives
\begin{align}
    \int_{0}^{2\pi} &\int_{0}^{2\pi} (\eta_{\beta} - \eta_{\alpha})^2 \cos(2\beta - 2 \alpha) \, d \alpha d \beta \nonumber \\ &= -2 \int_{0}^{2\pi} \int_{0}^{2\pi} \eta_{\beta} \eta_{\alpha} \cos(2\beta - 2 \alpha) \, d \alpha d \beta .
    \label{eq43}
\end{align}
$\alpha$ and $\beta$ integrals further separate when we expand the term $\cos(2\beta - 2 \alpha)$:
\begin{align}
    &\int_{0}^{2\pi} \int_{0}^{2\pi} \eta_{\beta} \eta_{\alpha} \cos(2\beta - 2 \alpha) \, d \alpha d \beta \nonumber \\ &= \int_{0}^{2\pi} \int_{0}^{2\pi} \eta_{\beta} \eta_{\alpha} (\cos{2\alpha} \cos{2\beta} + \sin2\alpha \sin2\beta) \, d \alpha d \beta \nonumber \\ &= \Bigg[ \int_{0}^{2\pi} \eta_{\alpha} \cos2\alpha \, d\alpha\Bigg]^2 + \Bigg[ \int_{0}^{2\pi} \eta_{\alpha} \sin2\alpha \, d\alpha\Bigg]^2 \nonumber \\ &= \Bigg| \int_{0}^{2\pi} \eta_{\alpha} e^{\Tilde{i}2\alpha} \, d\alpha \Bigg|^2 = 4 \pi^2 \big|\hat{\eta}_{\alpha}(-2)\big|^2 ,
    \label{eq44}
\end{align}
where $\hat{\eta}_{\alpha}$ is the Fourier transform of $\eta_{\alpha}$ defined by
\begin{equation}
    \hat{\eta_{\alpha}}(m) = \frac{1}{2\pi} \int_{0}^{2\pi} \eta_{\alpha} e^{-\Tilde{i} m \alpha} \, d\alpha.
\end{equation}
We study this in the Hilbert space $L^2(\mathbb{S}^1)$ with the inner product,
\begin{equation}
    \mu_{\alpha} \cdot \nu_{\alpha} = \frac{1}{2\pi} \int_{0}^{2\pi} \mu_{\alpha} \nu_{\alpha} \, d\alpha .
\end{equation}
Let $\mu_{\alpha} = \cos2\alpha$, $\nu_{\alpha} = \sin2\alpha$. Then we have $||\mu_{\alpha}||^2 = ||\nu_{\alpha}||^2 = 1/2$ and $\mu_{\alpha} \cdot \nu_{\alpha} = 0$. We express $\eta_{\alpha}$ as a linear combination of $\mu_{\alpha}$, $\nu_{\alpha}$, and a function $\eta_{\alpha}^{\perp}$ (orthogonal to both $\mu_{\alpha}$ and $\nu_{\alpha}$) as:
\begin{equation}
    \eta_{\alpha} = p \frac{\mu_{\alpha}}{||\mu_{\alpha}||} + q \frac{\nu_{\alpha}}{||\nu_{\alpha}||} + \eta_{\alpha}^{\perp},
\end{equation}
where $\mu_{\alpha} \cdot \eta_{\alpha}^{\perp} = \nu_{\alpha} \cdot \eta_{\alpha}^{\perp} = 0$. Moreover, we get
\begin{equation}
    ||\eta_{\alpha}||^2 = p^2 + q^2 + ||\eta_{\alpha}^{\perp}||^2.
    \label{eq48}
\end{equation} 
Now it is easy to see that
\begin{align}
    \big|\hat{\eta}_{\alpha}(-2)\big|^2 &= \Bigg| \frac{1}{2\pi} \int_{0}^{2\pi} \eta_{\alpha} e^{\Tilde{i} 2 \alpha} \, d\alpha \Bigg|^2 \nonumber \\ &= \Bigg[ \int_{0}^{2\pi} \eta_{\alpha} \cos2\alpha \, d\alpha\Bigg]^2 + \Bigg[ \int_{0}^{2\pi} \eta_{\alpha} \sin2\alpha \, d\alpha \Bigg]^2 \nonumber \\ &= (\eta_{\alpha} \cdot \mu_{\alpha})^2 + (\eta_{\alpha} \cdot \nu_{\alpha})^2 \nonumber \\
    &= (p||\mu_{\alpha}||)^2 + (q||\nu_{\alpha}||)^2 = \frac{p^2+q^2}{2}.
    \label{eq49}
\end{align}

Finally using Eqs.~\eqref{eq43},~\eqref{eq44},~\eqref{eq48}, and~\eqref{eq49}, from Eq.~\eqref{sec-var} we get
\begin{align}
    &\Gamma(\eta) \nonumber \\ &= 4\pi \Bigg[ \sqrt{1-E^2}(p^2+q^2+||\eta_{\alpha}^{\perp}||^2) - K\Big(\frac{p^2+q^2}{2}\Big) \Bigg] \nonumber \\ &= 4\pi \Bigg[ (p^2+q^2) \Big( \sqrt{1-E^2} - \frac{K}{2}\Big) + \sqrt{1-E^2} ||\eta_{\alpha}^{\perp}||^2 \Bigg].
    \label{eq50}
\end{align}
So, $\Gamma$ is positive definite when $\sqrt{1-E^2} - K/2 >0$. This implies the stability boundary is
\begin{align}
    E_T(K) =  \sqrt{1 - \frac{K^2}{4}} . \label{critical}
\end{align}
When  $E = 0$ we recover the critical coupling $K_c = 2$. Figure~\ref{bif-diagram} shows this agrees with simulation; the thick black curve on RHS of the lopsided bell correctly demarcates the phase wave stability region (cyan). To the north is the unsteady region, while to the east lies the sync state; we go east first.
\begin{figure}[t!]
\centering
\includegraphics[width = \columnwidth]{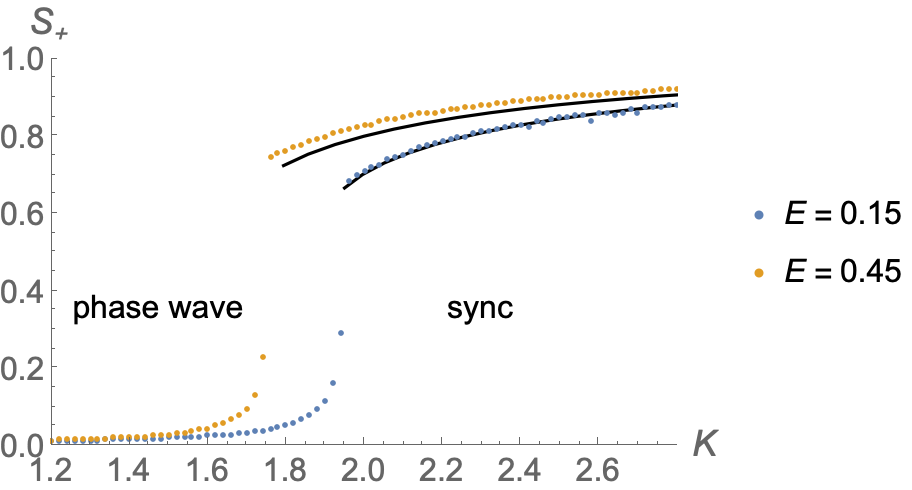}
\caption{ \textbf{Order parameter $S_+(K)$ for different driving strengths $E$}. Colored dots show simulation results, thick black curves numerical solution of Eqs.~\eqref{q1},~\eqref{q2}. Simulation parameters: $(dt,T,N) = (0.25, 200, 200)$ and each are averaged over last 10 \% of data collected.}
\label{Sp-K-e}
\end{figure}

%%%%%%%%%%%%%%%%%%%%%%%%%%%%%%%%%%%%%%%%%%%%%%%%%%%%%%%%%%%%%%%%%%%%%%%%%%%%%%%%%%%%
\textbf{Sync state}. The analysis carries through as before. In presence of driving, Eqs.~\eqref{s1} and~\eqref{s2} become
\begin{align}
    & 2 E + 2 \sin \Big( \frac{\xi}{2} - \alpha \Big) + K S_+ \sin \xi  = 0 \label{q1} , \\
    & S_+  = \frac{1}{2\pi}\int_{0}^{2\pi} e^{\Tilde{i} \xi(\alpha)} d \alpha \label{q2}.
\end{align}
Figure~\ref{Sp-K-e} shows the numerical solution to these matches simulation for different values of $E$. Now we present the last piece of our analysis: the $ K > 0, E \gg 0$ regime where collective behavior is unsteady.
\begin{figure*}[htpb!]
\centering
\includegraphics[width = 0.85 \textwidth]{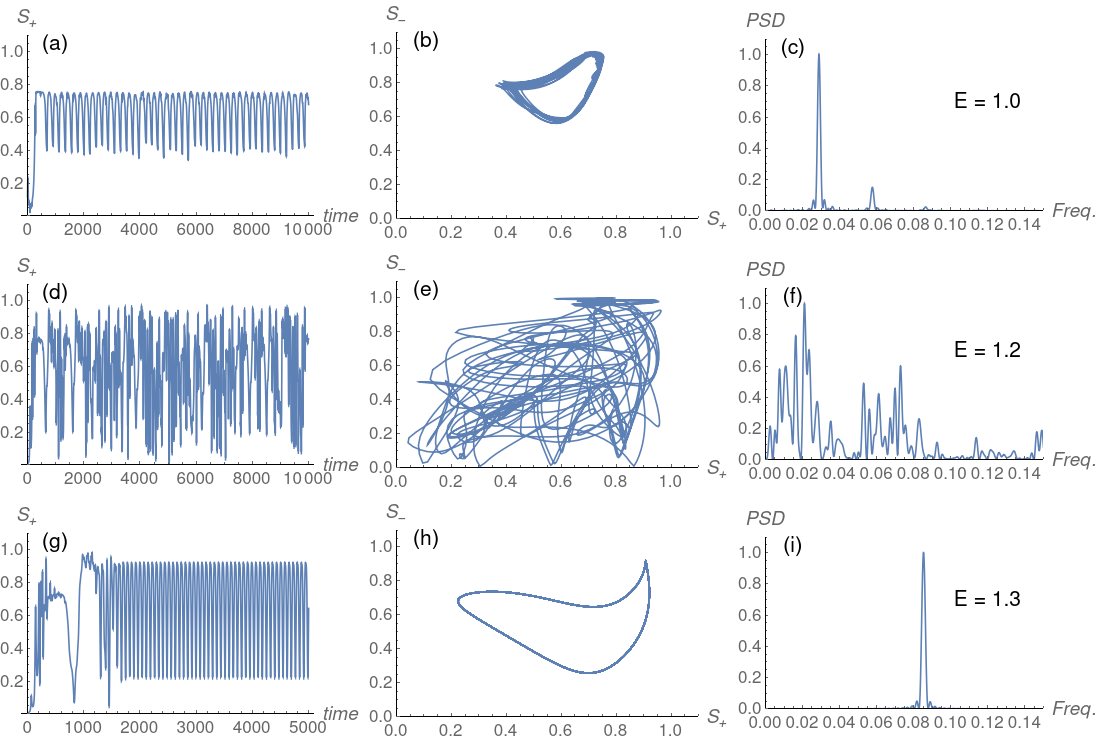}
\caption{{\bf Transition to chaos}. Time series of $S_+$, phase space trajectory in the $S_+$-$S_-$ plane, and the normalized power spectrum for different $E$ values when $K=0.5$. In the top row, $E=1.0$ which is the  quasi-periodic state. In the middle row, $E=1.2$ which is the chaotic state. In the bottom row, $E=1.3$ which is the  periodic orbit.}
\label{psd}
\end{figure*}

\textbf{Chaos \& quasi-periodicity}. Starting in the phase wave and moving in the $+E$ direction, we encounter low dimensional chaos (where the order parameters $W_{\pm}$ move chaotically in the complex plane). Figure~\ref{psd} explores the transition for $K  = 0.5$ (for which the critical driving is $E_T = \sqrt{15} / 4 \approx 0.968$). For $E = 1.0$, just past the threshold, $S_+$ is quasi-periodic. The time series is irregular, plots of $(S_+, S_-)$ form space-filling curves, and its power spectrum has a few dominant peaks with multiple smaller peaks (Fig.~\ref{psd} top row). As $E$ is increased to $E = 1.2$, the motion becomes chaotic. The time series and $(S_-, S_+)$ plots are irregular while the power spectrum has a wide band of frequencies (Fig.~\ref{psd} middle row). As $E$ increases to $1.3$ the chaos dies and simple periodic behavior is observed as indicated by the 1D manifold in $(S_-, S_+)$ plane and a single peak in the Fourier spectrum (Fig.~\ref{psd} bottom row). Here the swarmalators form traveling loops in $(x,\theta)$ space like ``flying saucers" (Supplementary Movie 1\&2). As $E$ increases, the width of these loops shrinks and finally become delta point masses at $E \rightarrow \infty$. 

%$S_{\pm}$ evolve periodically and attain both high (close to 1) and low (close to 0) values, indicating 'yoyo-ing' between synchrony and incoherence. Which means on the ring the swarmalators tend to synchrony at some time and moments after that they become desynchronized and the pattern repeats in a periodic manner. This simple periodic behavior exists for large $E$ and $K > 0$. In it, the swarmalators form traveling loops in $(x,\theta)$ space like ``flying saucers" (Supplementary Movie 1\&2). As $E$ increases, the width of these loops shrinks and finally become delta point masses at $E \rightarrow \infty$. 

\begin{figure}[htpb!]
\centering
\includegraphics[width = 1 \columnwidth]{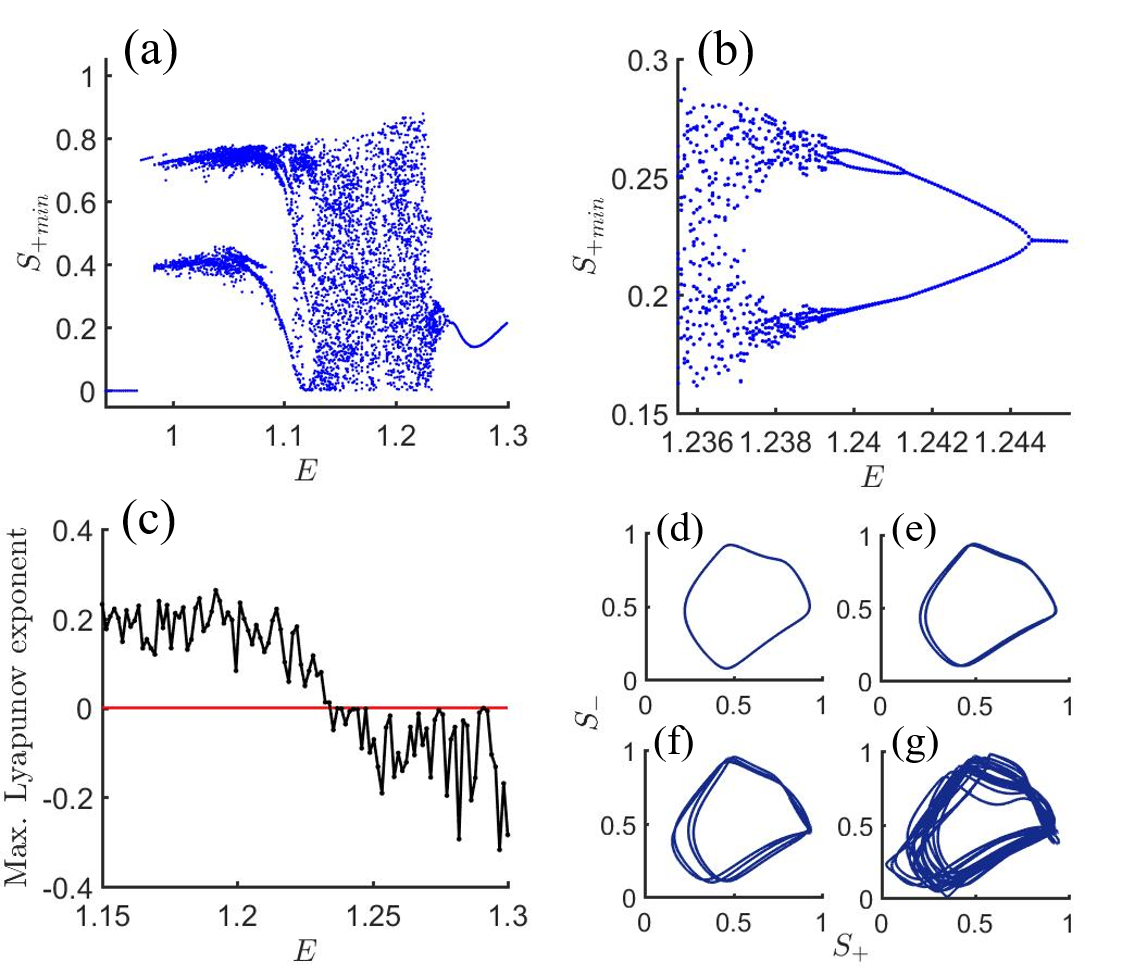}
\caption{{\bf Route to chaos}. (a) Bifurcations structure of $S_+(E)$. (b) Zoom in of period doubling route to chaos. (c) Lyapunov exponents. (d)-(g) Period doubling illustrated in the $(S_+, S_-)$ plane. $E = 1.248, 1.243, 1.241, 1.23$ for the panels (d), (e), (f), and (g), respectively. In all panels, $K = 0.5$ and $(dt,T,N) = (0.01,10000,200)$. Last 20 \% data were taken to calculate peaks of the time series of $S_+$.}
\label{chaos}
\end{figure}

\begin{figure}[t]
\centering
\includegraphics[width =0.85 \columnwidth]{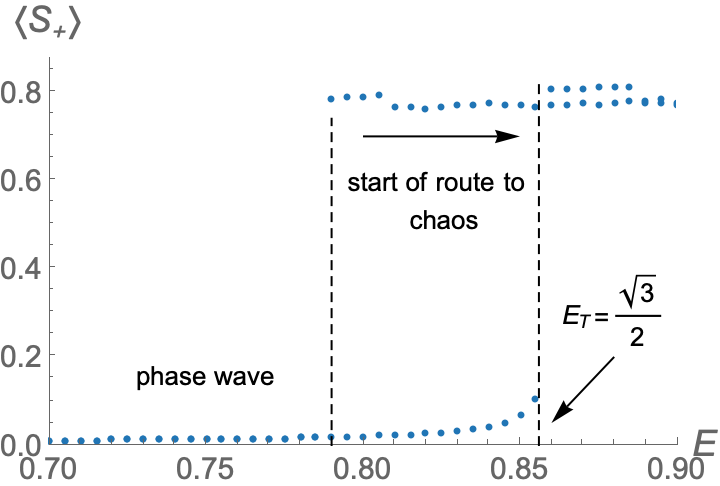}
\caption{\textbf{Hysteresis} in the time average of order parameter $\langle S_+ \rangle$ versus $E$ for $K = 1$ (for which $E_T = \sqrt{3} / 2$ from the analytical expression Eq.~\eqref{critical}). Lower branch, phase wave state. Upper branch,  start of route to chaos. Simulation parameters: $(dt,T,N) = (0.25, 100, 200)$.  Each data point is the time average of the last $10 \%$ of data. }
\label{hysteresis}
\end{figure}

Figure~\ref{chaos} further explores the route to chaotic region. Panel (a) shows the bifurcation structure of $\min S_+(K)$. Starting from $E \approx 1.3$ and \textit{decreasing} $E$, the classic period doubling route to chaos is observed (panel (b) zooms in the period doubling region). Panel (c) shows the maximum Lyapunov exponents are consistent with this picture: at the onset of chaos $E \approx 1.24$, they switch from negative to positive. Lastly, panels (d)-(g) illustrate the period doubling in the $(S_-, S_+)$ plane. Starting from the right at $E=1.248$, where 1-periodic orbit exists (Fig.~\ref{chaos} (d)), one can observe period doubling cascade with decreasing $E$. At $E=1.243$, a 2-periodic orbit (Fig.~\ref{chaos} (e)) and at $E=1.241$, a 4-periodic orbit (Fig.~\ref{chaos} (f)) are found before the motion becomes chaotic (Fig.~\ref{chaos} (g) for $E=1.23$). Supplementary Movie 3 shows the $(S_+, S_-)$ plots along with accompanying scatter plots in the $(x, \theta)$. Finally, Fig.~\ref{hysteresis} shows the transition to unsteady behavior is hysteretic. For $K=1$, the phase wave loses stability at $E_T = \sqrt{3}/2 \approx 0.87$. A quasi-periodic orbit is born beyond this value of $E$. However, the backward transition shows that quasi-periodicity is sustained till $E \approx 0.79 < E_T$.

\section{Robustness to non-symmetric pinning}
Here we briefly demonstrate that our results are robust to small perturbation around the symmetric pinning idealization $\alpha_i = \beta_i$. We explore two choices: symmetric pinning with an offset (i) $\alpha_i = \beta_{i+1} = 2\pi i / N$ and (ii) randomly drawn (but sorted) pinning $\alpha_i, \beta_i \sim U(0,2\pi)$. 
\begin{figure}[t]
\centering
\includegraphics[width = 0.9 \columnwidth]{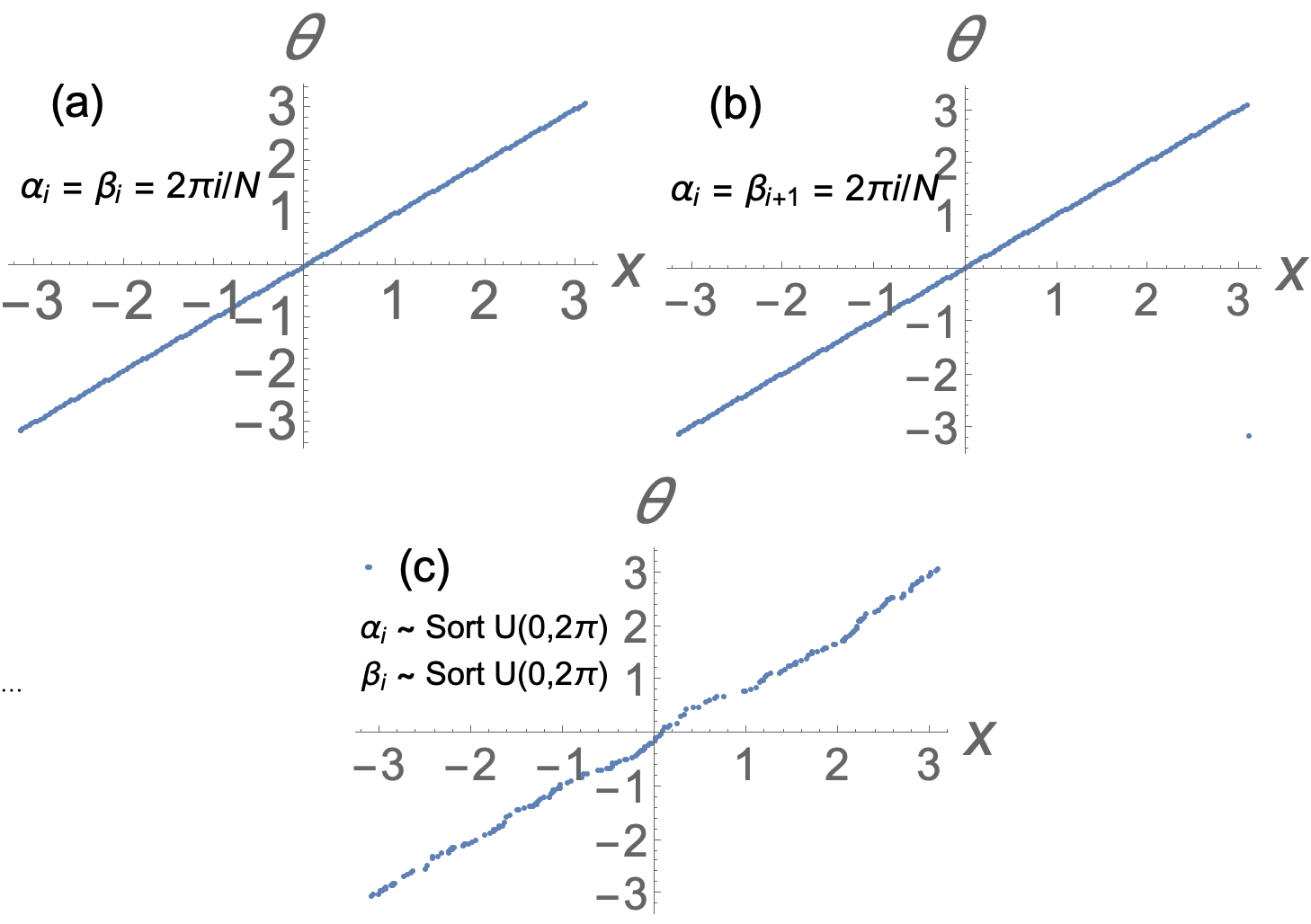}
\caption{\textbf{Robustness to asymmetric pinning}. The phase wave persists under small amounts of disorder around the linearly spaced model. The other collective states are deformed in a qualitative manner. Here $(N,dt,T) = (300,0.25,100)$.}
\label{robustness}
\end{figure}
We find each collective state persists for both (i) and (ii). Figure~\ref{robustness} shows the phase wave as an example. We also computed the numerical bifurcation diagram $(K,E)$ and found it was qualitatively similar. We leave further explorations for future work.

%%%%%%%%%%%%%%%%%%%%%%%%%%%%%%%%%%%%%%%%%%%%%%%%%%%%%%%%%%%%%%%%%%%%%%%%%%%%%%%%%%%%%%%%%%%%%%%
\section{Discussion}
Our hope is that the 1D swarmalator model be used as toy model for general studies of swarmalation in disordered media. It seems to have the right ingredients: it is tractable yet also mimics real world behavior. The phase wave state is observed in many 1D swarmalators, such as circularly confined spermatazoa \cite{creppy2016symmetry} and bordertaxic vinegar eels \cite{quillen2021metachronal,quillen2022fluid,peshkov2022synchronized}. The split phase wave mimics the `wavy-antisynchronization' seen in  models of Japanese tree frogs (see Fig.3 in \cite{aihara2014spatio}) and the chain states of Janus matchsticks (Fig 5(c) of \cite{chaudhary2014reconfigurable}). As for the dynamic states, chaos has been observed in the dynamics of single magnetic domains walls \cite{chaos_mag_wall}, but to our knowledge has not yet been observed in the $N > 1$ regime. On the other hand, so-called biological turbulence is often observed in aggregations of microswimmers \cite{PhysRevLett.110.228102}. Could the material defects in the host medium be the source of this unsteady behavior? If so, our model could provide a tractable setting to examine this form of chaos (indeed, to our knowledge, this is the only oscillator model with pinning that produces chaos).

To sum up, the model's partial matches to reality are encouraging, but it's still not clear if the model is general enough for broad use. The radical idealizations we made (mean field coupling, symmetric pinning $\alpha_i = \beta_i$) likely cut out some essential physics. Nevertheless, we hope the model at least finds use as an early stepping stone in the study of pinning in systems which both sync and swarm -- systems for which theories are currently lacking.

Ambitious follow up work could try to derive the evolution equations for the order parameters $\dot{S}_{\pm}$. Crawford did this for the Kuramoto model via a center manifold calculation \cite{crawford1994amplitude}. If we could adapt his analysis for $\dot{S}_{\pm}$ we could crack the bifurcations of the unsteady states. For instance, we could study the tricritical point joining the phase wave, sync, and unsteady phases (the point at which the red, cyan and magenta region meet in Figure~\ref{bif-diagram}). Future work could also relax our model's symmetry. Instead of a common $(E,b,K)$ appearing in both the $\dot{x}, \dot{\theta}$, one could split them up $E \rightarrow (E_x, E_{\theta}), b \rightarrow (b_x, b_{\theta}), K \rightarrow (K_x, K_{\theta})$. Local coupling, non-symmetric pinning $\alpha_i \neq \beta_i$, and spatial motion in 2D would also be interesting to explore. 

%Future work could relax our model's symmetry. Instead of a common $(E,b,K)$ appearing in both the $\dot{x}, \dot{\theta}$ equations, one could split them into $E \rightarrow (E_x, E_{\theta}), b \rightarrow (b_x, b_{\theta}), K \rightarrow (K_x, K_{\theta})$. Local coupling, non-symmetric pinning $\alpha_i \neq \beta_i$, and spatial motion in 2D would also be interesting to explore. 

%Ambitious follow up research could try to derive the evolution equations for the order parameters, $\dot{S}_{\pm}$.  Crawford did this for the regular Kuramoto model via a center manifold calculation \cite{crawford1994amplitude}. Adapting this to swarmalators would be a serious undertaking. But if we could do it, we could crack the bifurcations in the unsteady regime. For instance, we could study the tricritical point joining the phase wave, sync, and unsteady phases (the point at which the red, cyan and magenta region meet in Figure~\ref{bif-diagram}). Another future direction would be to relax our model's extreme symmetry. Instead of a common $(E,b,K)$ appearing in both the $\dot{x}, \dot{\theta}$, one could split them into $(x, \theta)$ components: $E \rightarrow (E_x, E_{\theta}), b \rightarrow (b_x, b_{\theta}), K \rightarrow (K_x, K_{\theta})$. Disorder other than the symmetric $\alpha_i = \beta_i$ would also be interesting to explore. Finally, the model could be adapted to the more realistic case where swarmalators are free to move in 2D. 

%%%%%%%%%%%%%%%%%%%%%%%%%%%%%
\textbf{Reproducibility}. Code used for simulations and analytic calculations available on Github \footnote{https://github.com/Khev/swarmalators/tree/master/1D/on-ring/random-pinning}.

%%%%%%%%%%%%%%%%%%%%%%%%%%%%%%%%%%%%%%%%%%%%%%%%%%%%%%%
    
\bibliographystyle{apsrev}
\bibliography{ref.bib}

\end{document}